\documentclass[aps,prl,twocolumn,noshowpacs,superscriptaddress,groupedaddress, floatfix,superscriptaddress, preprintnumbers]{revtex4-1}
 
\usepackage{graphicx}
\usepackage{dcolumn}
\usepackage{bm}
\usepackage{amsmath, amssymb}
\usepackage[breaklinks,colorlinks=true]{hyperref}

\usepackage{multirow}
\usepackage{feynmf}
\usepackage{slashed}
\usepackage{epstopdf}

\usepackage[usenames]{color}

\usepackage{float}
\usepackage{soul}

\newcommand{\be}{\begin{equation}}
\newcommand{\ee}{\end{equation}}
\newcommand{\ba}{\begin{eqnarray}}
\newcommand{\ea}{\end{eqnarray}}

\newcommand{\nta}{\Delta} 
\newcommand{\tam}{\Theta} 
\newcommand{\tae}{\Theta} 

\newcommand{\LambdaQCD}{\Lambda_{\text{QCD}}}
\newcommand{\nB}{n_{\mathrm{B}}}
\newcommand{\muB}{\mu_{\mathrm{B}}}
\newcommand{\Tc}{T_{\mathrm{c}}}
\newcommand{\Nc}{N_{\mathrm{c}}}

\begin{document}

\begin{flushright}
\end{flushright}

\preprint{INT-PUB-22-019}

\title{Trace anomaly as signature of conformality in neutron stars}
\author{Yuki~Fujimoto}
\email{yfuji@uw.edu}
\affiliation{Institute for Nuclear Theory, University of Washington, Box 351550, Seattle, WA, 98195, USA}

\author{Kenji~Fukushima}
\email{fuku@nt.phys.s.u-tokyo.ac.jp}
\affiliation{Department of Physics, The University of Tokyo, 7-3-1 Hongo, Bunkyo-ku, Tokyo 113-0033, Japan}

\author{Larry~D.~McLerran}
\email{mclerran@me.com}
\affiliation{Institute for Nuclear Theory, University of Washington, Box 351550, Seattle, WA, 98195, USA}

\author{Micha\l~Prasza\l owicz}
\email{michal.praszalowicz@uj.edu.pl}
\affiliation{Institute of Theoretical Physics, Jagiellonian University, S. \L ojasiewicza 11, 30-348 Krak\'ow, Poland}
\affiliation{Institute for Nuclear Theory, University of Washington, Box 351550, Seattle, WA, 98195, USA}

\date{\today}

\begin{abstract}
  We discuss an interpretation that a peak in the sound velocity in
  neutron star matter, as suggested by the observational data,
  signifies strongly-coupled conformal matter.
  The normalized trace anomaly is a dimensionless measure of
  conformality leading to the derivative and the non-derivative
  contributions to the sound velocity.
  We find that the peak in the sound velocity is attributed to the
  derivative contribution from the trace anomaly that steeply
  approaches the conformal limit.
  Smooth continuity to the behavior of high-density QCD implies that
  the matter part of the trace anomaly may be positive definite.
  We discuss a possible
  implication of the positivity condition of the trace anomaly on the
  $M$-$R$ relation of the neutron stars.
\end{abstract}

\maketitle

\paragraph*{Introduction:}

Massless quantum chromodynamics (QCD) exhibits conformal
symmetry, and the expectation value of the trace of the energy-momentum tensor,
$\langle \tam\rangle \equiv \langle T^{\mu}_{\;\;\mu} \rangle$, 
vanishes at the classical level~\footnote{For the
  non-Abelian gauge theories coupled to fermions scale invariance
  implies conformality; see Ref.~\cite{Polchinski:1982qk}.  Throughout
  this paper \textit{conformality} means
  $\langle\tam\rangle=0$. Conversely, we dub
  $\langle\tam\rangle\neq 0$ as a \textit{trace anomaly}.}.\nocite{Polchinski:1982qk}
Conformal symmetry, however, is broken at the quantum level.
This violation is quantified via the trace anomaly, which has the
anomalous term proportional to the gluon condensate owing to the
running of the strong coupling constant, $\alpha_s$.

At finite temperature $T$ and baryon
chemical potential $\muB$, the condensate should depend on $T$ and
$\muB$ and we can decompose the trace anomaly into the vacuum and
the matter parts.
The matter part of the trace anomaly can be expressed in
terms of thermodynamic quantities, i.e.,
the energy density $\varepsilon$ and the pressure $P$,
as $\langle\tam\rangle_{T,\muB} = \varepsilon-3P$.
An interesting question is how $\langle\tae\rangle_{T,\muB}$ changes
near the transition point.
At finite $T$ and $\muB/T\ll 1$ the lattice-QCD simulations provide
the first-principles estimate.
In Refs.~\cite{Boyd:1995zg,Boyd:1996bx} the normalized trace anomaly,
$(\varepsilon-3P)/T^4$
(referred to as the interaction measure), in the pure
Yang-Mills theory was found to have
a sharp peak at the deconfinement temperature, $\Tc$,
and a tail approaching zero asymptotically at high $T$.

This enhancement is understood from the thermal
modification of the condensate.  The gluon condensate melts near the
transition point leading to a peak in the thermal part of the trace
anomaly.
Lattice measurements of the trace anomaly have a striking impact on
our understanding of deconfined matter.
As pointed out in the section of ``Discussion of Conformal Symmetry''
in Ref.~\cite{Miller:2006hr} the trace
anomaly behaves like $\langle\tae\rangle_T \propto T$ even for
$T\gtrsim 2\Tc$ suggesting that a strongly-coupled gluonic system is
realized in the deconfined phase.

The trace anomaly has been also calculated in full QCD 
with dynamical quarks (e.g., Refs.~\cite{Cheng:2007jq,
Borsanyi:2013bia,  HotQCD:2014kol}).
The hard-thermal-loop perturbation theory (HTLpt) is successful
in reproducing the trace anomaly with quarks already around $T\sim 2\Tc$,
while the agreement between the lattice and the HTLpt results
for the pure Yang-Mills theory begins only around
$T\sim 8\Tc$~\cite{Andersen:2011ug}.

These high-$T$ studies motivate us to investigate the trace anomaly at
high baryon density.
For baryon density $\nB > n_0$, where
$n_0 \approx 0.16\,\text{fm}^{-3}$ is the saturation density, 
QCD thermodynamics is elusive because the lattice calculations are
hampered by the sign problem.
The only \textit{ab initio} methods are: the chiral
effective field theory ($\chi$EFT) around $\nB\sim n_0$
(see, e.g., Ref.~\cite{Drischler:2021kxf} for a recent review), and
the perturbative QCD (pQCD) at high density where
$\alpha_s$ is sufficiently small~\cite{Freedman:1976xs,
  *Freedman:1976dm, *Freedman:1976ub, Kurkela:2009gj} (see also
Refs.~\cite{Gorda:2018gpy, Gorda:2021znl,*Gorda:2021kme,
  Gorda:2022fci,*Gorda:2022zyc, Fujimoto:2020tjc, Fernandez:2021jfr}
for recent developments).

To constrain thermodynamic quantities or the equation of state
(EoS), we can also rely on the empirical
knowledge from the neutron star (NS) observations;
the sound velocity, $v_s^2 \equiv dP/d\varepsilon$,
characterizes the EoS\@.
Recently, a non-monotonicity of $v_s^2$ as a function of
density has been conjectured~\cite{Kojo:2020krb,Altiparmak:2022bke,Ecker:2022xxj}.
For instance, Quarkyonic description of dense
matter~\cite{McLerran:2007qj,Duarte:2021tsx,Kojo:2021ugu,Kojo:2021hqh,Fukushima:2015bda,McLerran:2018hbz,Jeong:2019lhv,Sen:2020peq,Cao:2020byn,Kovensky:2020xif}
in the large-$N_c$ limit~\cite{tHooft:1973alw,Witten:1979kh}
leads to the rapid increase, accompanied by a peak
of the sound velocity (see also
Refs.~\cite{Pisarski:2021aoz,Hippert:2021gfs,Lee:2021hrw,Marczenko:2022jhl}).

At asymptotic densities where QCD recovers conformality,
$v_s^2 \to 1/3$ is expected;  this limit is
commonly referred to as the conformal limit, and thus $1/3-v_s^2$ serves a
measure of conformality.
There was a conjecture claiming $1/3-v_s^2\ge 0$ at all
densities~\cite{Cherman:2009tw}; see also Ref.~\cite{Hohler:2009tv}.
However, the recent analyses of NS data including
the two-solar-mass
pulsars~\cite{Demorest:2010bx,*Fonseca:2016tux, Antoniadis:2013pzd,
  NANOGrav:2019jur,*Fonseca:2021wxt,Romani:2022jhd}
are in strong tension with $1/3-v_s^2\ge 0$ at sufficiently high
$\nB$~\cite{Bedaque:2014sqa,Tews:2018kmu, Fujimoto:2017cdo, *Fujimoto:2019hxv,
  *Fujimoto:2021zas, Drischler:2020fvz,Drischler:2021bup}, which
seems to challenge the conformality in dense NS matter in deep cores.

Here we propose the trace anomaly scaled by
the energy density as a new measure of conformality. 
The sound velocity is expressed solely in terms of the normalized
trace anomaly, and the latter
is a more comprehensive quantity than $v_s^2$.
Here, we extract the trace anomaly from the EoSs inferred from the
NS data~\cite{Fujimoto:2017cdo, *Fujimoto:2019hxv,
  *Fujimoto:2021zas, Al-Mamun:2020vzu, Raaijmakers:2021uju, Gorda:2022jvk}.
We discuss the conformal limits $\langle\tam\rangle_{T,\muB}\to 0$ and
$v_s^2\to 1/3$, and clarify the difference.
We show that the enhancement in the sound velocity is not in 
contradiction with conformality.
We then discuss the possibility that the trace anomaly is positive
definite at all densities.  We give a number of arguments for the
positivity of the trace anomaly and discuss implications for NS physics.
\vspace{0.5em}

\paragraph{Trace anomaly at finite baryon density:}

Scale transformations lead to the dilatation current $j_D^\nu = x_\mu
T^{\mu\nu}$ for which
$\partial_\nu j_D^\nu = T^{\mu}_{\;\;\mu}=\tam$~\cite{Coleman:1985rnk}.
For conformal theories $\tam=0$ but in QCD both quark masses and the trace
anomaly explicitly break the scale invariance as~\cite{Collins:1976yq,Nielsen:1977sy}
\begin{equation}
  \label{eq:ta}
  \tam = \frac{\beta}{2g} F_{\mu\nu}^a F^{\mu\nu}_a
  + (1+\gamma_m) \sum_f m_f \bar{q}_f q_f \,,
\end{equation}
where
$\beta/2g=-(11-2N_f/3)\alpha_s/8\pi+\mathcal{O}(\alpha_s^2)$ is the
QCD beta function and $\gamma_m=2\alpha_s/\pi+\mathcal{O}(\alpha_s^2)$
is the anomalous dimension of the quark mass.

At finite $T$ and/or $\muB$, the expectation value involves a matter
contribution as
$\langle \tae \rangle = \langle \tae \rangle_{T,\muB} + \langle \tae \rangle_0$
where $\langle \tae\rangle_0$ represents the vacuum expectation
value at $T=\muB=0$.
In this work we will focus on the matter contribution only given by
\begin{equation}
  \langle \tae \rangle_{T,\muB} = \varepsilon - 3P\,.
\end{equation}
It is customary to call $\langle \tae\rangle_{T,\muB}$ the trace
anomaly too.
If thermal degrees of freedom are dominated by massless particles
as is the case in the high-$T$ limit, the Stefan-Boltzmann law is saturated and $P\sim T^4$ at high
temperature or $P\sim \muB^4$ at high density, so that
$\varepsilon=3P$.
Conversely, using thermodynamic relations, one can show that
$\langle\tae\rangle_{T,\muB}= 0$ implies $P\propto T^4$ or $P\propto
\muB^4$, respectively.
Thus, $\langle\tae\rangle_{T,\muB}$ is a probe for the thermodynamic
content of matter.

The physical meaning of the trace anomaly is transparent from the
following relations:
\begin{equation}
  \label{eq:tracedof}
  \frac{\langle\tae\rangle_{T,\muB=0}}{T^4} = T \frac{d\nu_T}{dT} \,,\qquad 
  \frac{\langle\tae\rangle_{T=0,\muB}}{\muB^4} = \muB 
  \frac{d\nu_\mu}{d\muB}\,,
\end{equation}
where we quantify the effective degrees freedom by
$\nu_T \equiv P/T^4$ and
$\nu_\mu \equiv P/\muB^4$
for hot matter at $\muB=0$ and dense matter at $T=0$, respectively.
These imply that the trace anomaly is proportional to the increasing rate of
the thermal degrees of freedom as the temperature/density grows up.

Here, we propose to use
\begin{equation}
  \nta \equiv \frac{\langle\tae\rangle_{T,\muB}}{3\varepsilon} = \frac13 -\frac{P}{\varepsilon}\,.
\end{equation}
as a measure of the trace anomaly~\footnote{Our $\nta$ is equivalent to $\mathcal{C}$ defined
  in Ref.~\cite{Gavai:2004se} apart from an overall constant, $1/3$.}.
The thermodynamic stability and the causality require $P>0$ and $P \le
\varepsilon$, respectively.
Therefore $-2/3\le \nta <1/3$, and
$\nta \to 0$ in the scale-invariant limit.

We decompose the sound velocity as
\begin{equation}
  \label{eq:vs2}
  v_s^2 = \frac{dP}{d\varepsilon} = v_{s, \,\text{deriv}}^2 +
  v_{s,\, \text{non-deriv}}^2\,,
\end{equation}
where the derivative and the non-derivative terms are
\begin{equation}
  \label{eq:vs3}
  v_{s,\,\text{deriv}}^2 \equiv -\frac{d\nta}{d\eta}\,,\qquad
  v_{s,\,\text{non-deriv}}^2 \equiv \frac{1}{3} -\nta\,.
\end{equation}
Here, $\eta\equiv \ln(\varepsilon/\varepsilon_0)$
and $\varepsilon_0$ is
the energy density at nuclear saturation density, i.e.,
$\varepsilon_0=150\,\mathrm{MeV}/\mathrm{fm}^3$.
From these expressions it is evident that the restoration of
conformality renders $\nta\to 0$ and $d\nta/d\eta\to 0$, so that
$v_s^2\simeq v_{s,\,\text{non-deriv}}^2 \to 1/3$ in the conformal
limit at asymptotically high density.
\vspace{0.5em}

\paragraph*{Trace anomaly from the NS observations:}

\begin{figure}
  \includegraphics[width=\columnwidth]{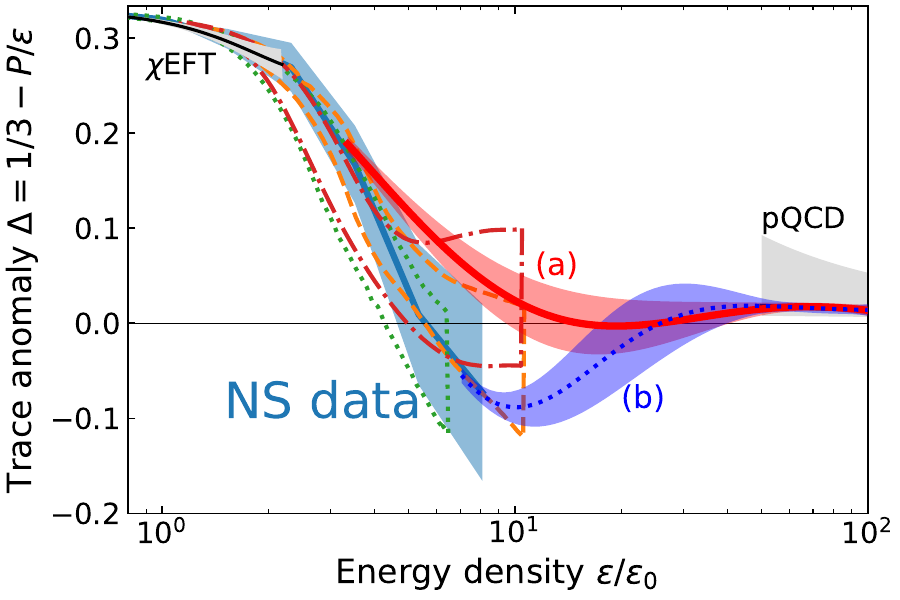}
  \caption{Normalized trace anomaly read out from four independent
    EoSs inferred from NS
    data; the light blue solid line and error band from
    Ref.~\cite{Fujimoto:2017cdo,*Fujimoto:2019hxv,*Fujimoto:2021zas},
    the orange dashed lines from Ref.~\cite{Al-Mamun:2020vzu}, the
    green dotted lines from Ref.~\cite{Raaijmakers:2021uju}, and the
    red dot-dashed lines from Ref.~\cite{Gorda:2022jvk}.
    We show two \textit{ab initio} calculations
    ($\chi$EFT~\cite{Drischler:2020fvz} and
      pQCD~\cite{Kurkela:2009gj})
    and the red line marked as (a) and the blue dotted line marked as
    (b) are interpolations with $1\sigma$ band by the Gaussian process applied
    to different regions of NS data.}
  \label{fig:NS}
\end{figure}

In Fig.~\ref{fig:NS}, we show $\nta$
extracted from various $P(\varepsilon)$ constrained by NS
observables~\cite{Fujimoto:2017cdo,*Fujimoto:2019hxv,*Fujimoto:2021zas,
  Al-Mamun:2020vzu, Raaijmakers:2021uju, Gorda:2022jvk}.
The error band represents the $1\sigma$ credible interval
corresponding to the error in $P(\varepsilon)$.
Since $\varepsilon$ is treated as an explanatory variable,
the relative error in $\Delta(\varepsilon)$ is assumed to be the same
as that in $P(\varepsilon)$.

For all these data $\nta \sim 0$ within the error at relatively low
energy density.
Note that the red dash-dotted curve in Fig.~\ref{fig:NS} follows from
the analysis including pQCD as an input~\cite{Gorda:2022jvk},
which makes the tendency $\nta \sim 0$ more apparent.

\begin{figure}
  \includegraphics[width=0.48\textwidth]{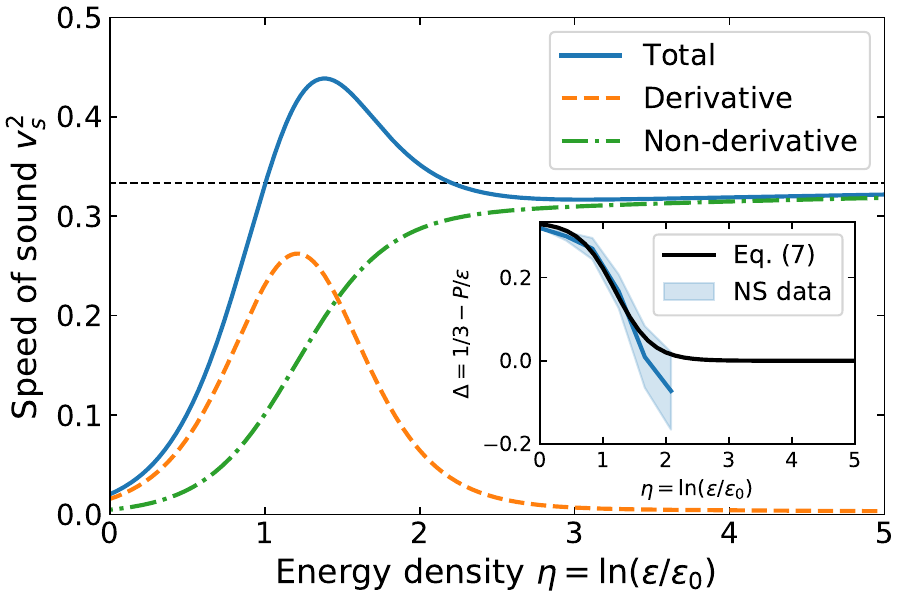}
  \caption{The speed of sound and its decomposition \eqref{eq:vs3}
    calculated from \eqref{eq:parametrized} as shown in the inset
    plot.  The horizontal axis is the logarithmic energy $\eta$
    normalized to the value at the saturation point, $\varepsilon_0 =
    150\,\text{MeV}/\text{fm}^3$.}
  \label{fig:parametrized}
\end{figure}

Fig.~\ref{fig:NS} shows that the (normalized) trace anomaly
in the present experimental range
monotonically decreases with increasing $\varepsilon$.
At asymptotically high density $\Delta\to 0$ should be eventually
reached.
It is nontrivial that the NS observations favor $\Delta\sim 0$
at intermediate $\varepsilon$, well below the asymptotic density.
Here, we elucidate that this quick approach to conformality causes a
prominent peak in $v_s^2$.
We emphasize that, even if the behavior toward $\Delta \to 0$ is
monotonic in $\varepsilon$, $v_s^2 > 1/3$ can be induced.

The minimal parametrization of monotonically decreasing $\nta$ is
\begin{equation}
  \nta =  \frac13 - \frac13 \cdot \frac1{e^{-\kappa(\eta - \eta_c)} + 1}
  \left(1-\frac{A} {B +\eta^2}\right)\,.
  \label{eq:parametrized}
\end{equation}
The crossover density to conformal matter is characterized by
$\eta_c$ and the width of the crossover region is
$1/\kappa$.
Equation~\eqref{eq:parametrized}
has the correct limit
$\nta\to 1/3$ for $\eta \ll \eta_c$ and
$\nta \sim (A/3)/(B+\eta^2) \to 0$ for $\eta \gg \eta_c$.
Nonzero $A$ and $B$ represent the
pQCD logarithmic tails that are not well constrained from
the NS data.
One parameter set that fits the observational data reads,
\begin{equation}
  \kappa = 3.45, \quad \eta_c = 1.2\,\quad A = 2, \quad B = 20\,. 
\end{equation}
The fit together with data is shown in the inset plot in
Fig.~\ref{fig:parametrized}.
We show $v_s^2$ computed from
Eq.~\eqref{eq:vs2} with the help of Eq.~\eqref{eq:parametrized}
in Fig.~\ref{fig:parametrized}.
In the high density region for $\eta\gtrsim 2$,
$v_s^2$ is dominated by
$v_{s,\,\text{non-deriv}}^2$ approaching the conformal value, $1/3$.
At low density for $\eta\lesssim 1$,  $v_s^2$ goes to zero.

The most interesting is the behavior of $v_s^2$ around
$1\lesssim \eta\lesssim 2$.
This density region corresponds to the energy scale of
the transitional change from non-relativistic to relativistic degrees
of freedom.  There, $v_s^2$ develops a peak whose height can become
larger than the conformal value.

The dashed and the dash-dotted lines in Fig.~\ref{fig:parametrized}
show $v_{s,\text{deriv}}^2$ and $v_{s,\text{non-deriv}}^2$,
respectively.  Because $\nta$ of Eq.(\ref{eq:parametrized}) is a monotonic function,
$v_{s,\text{non-deriv}}^2$ smoothly increases with increasing $\eta$.
Thus, $v_{s,\text{deriv}}^2$ exhibits the peak structure.
From this decomposition we clearly recognize that the peak in $v_s^2$
is not caused by the violation of the conformal bound, but it is a
signature of the steep approach to the conformal limit!

We stress that this is quite different from high-$T$ QCD where the
normalized trace anomaly itself has a peak around $\Tc$, which causes a
\emph{minimum} in the sound velocity.  Along the $T$ axis conformality is
restored only at temperatures \emph{far above} $\Tc$.  One might have an
impression that conformality in QCD should be associated with the weak
coupling, but it is not necessarily the case.
What we find from Fig.~\ref{fig:NS} is
that conformality quantified by $\nta$ is quickly restored around
$1\lesssim \eta\lesssim 2$ and the peak in $v_s^2$ should be
interpreted as a signature of conformality.
The peak position may well be identified as the point of the slope
change as observed in Ref.~\cite{Annala:2019puf}.
Around this peak $\alpha_s$ is not yet small 
and the state of matter for $\eta\gtrsim 2$ should be regarded as
``strongly-coupled conformal matter''.

We note that $v_s^2\to 1/3$ generally occurs at lower density than $\nta\to 0$.
We can illustrate this in a simple model with the vector interaction
between the currents whose energy density is given by
\begin{equation}
  \label{eq:9}
  \varepsilon(n) = m_N \nB + \frac{C}{\Lambda^2} \nB^2\,, 
\end{equation}
where $m_N=\Nc \LambdaQCD$ is the baryon mass,
and $C$ and $\Lambda$ are
the typical interaction strength and the scale of the system,
respectively.
This can be thought of as the generalization of the mean-field quantum
hadrodynamics~\cite{Serot:1997xg}.
Note that
$\mu_{\text{B}} = m_N + 2(C/\Lambda^2)\nB$, and  
$P = (C / \Lambda^2) \nB^2$.
This means that $\tae = m_N \nB - 2(C/\Lambda^2) \nB^2$
and  $v_s^2 = 2(C/\Lambda^2)\nB / [m_N + 2(C/\Lambda^2)\nB]$.
The conformal point 
$\nta\to 0$ is
reached when $\nB \sim N_c \LambdaQCD^3/(2C)$.
The condition of $v_s^2\to 1/3$ is reached earlier at
$\nB \sim N_c \LambdaQCD^3 / (4C)$.
So in this model the density at which $v_s^2$ surpasses the conformal limit is
always lower than that for the trace anomaly.
\vspace{0.5em}

\paragraph*{Strongly-coupled conformal matter:}

In Fig.~\ref{fig:NS} we overlay the 
currently available \textit{ab initio} calculations of 
$\chi$EFT~\cite{Drischler:2020fvz} and pQCD~\cite{Kurkela:2009gj} on
the observational data that, however, do not constrain $\nta$ beyond
$\varepsilon/\varepsilon_0\sim 10^1$.
We utilized the Gaussian process for the interpolation using NS
data from the machine learning~\cite{Fujimoto:2019hxv} up to the
density $\varepsilon/\varepsilon_0\lesssim 4$ (a) and using all data
up to $\varepsilon/\varepsilon_0\sim 8$ (b).  Details about the
choice of the kernel and the noise will be reported elsewhere.

In the conservative inference in (a) $\nta$ stays positive or
slightly negative after quickly approaching zero, which implies
a possible bound, $\Delta \ge 0$.
Once the conformal limit of the trace anomaly is saturated,
the underlying theory becomes approximately scale invariant
and the EoS drastically simplifies.
Baryons are strongly interacting, and yet the resultant EoS of
strongly-coupled conformal matter is $P\approx \varepsilon/3$.

If the mean value from the machine learning inference is
extrapolated, the Gaussian process prefers (b).
In this case $\nta$ has a non-monotonic structure with two nodes.
Accordingly, there should be a density window with
$d \nta /d\varepsilon > 0$ (i.e., $v_{s,\text{(deriv)}}^2 < 0$)
between the two zeros.
The peak in $v_s^2$ is hardly affected, however, the maximum of
$v_s^2$ is pulled up as compared to (a).
If $v_{s,\text{(deriv)}}^2$ happens to be negative large, $v_s^2$
approaches zero after the peak, which causes softening of the
EoS similarly to the first-order phase transition.
Intuitively, the peak in $v_s^2$ is generated by EoS stiffening, but
the soft pQCD EoS at high density requires EoS softening at
intermediate density.
\vspace{0.5em}

\paragraph*{Is the trace anomaly positive in finite-density QCD?:}

Let us focus on the scenario (a) and consider its implications.
The smooth curve of (a) in Fig.~\ref{fig:NS} supports a
hypothetical relation,
$\langle \tam \rangle_{\muB} \geq 0$ (equivalently,
$P \leq \varepsilon /3$).
The positivity condition of the QCD trace anomaly has been often
assumed in the literature of finite-$T$ QCD; see, e.g.,
Ref.~\cite{Bjorken:1982qr}.
The lattice-QCD calculations at finite $T$ give thermodynamic
quantities satisfying
$\langle \tam \rangle_T \geq 0$~\cite{Cheng:2007jq, Borsanyi:2013bia, HotQCD:2014kol}.

In general, however, the trace anomaly may not be positive definite.
For example if the low-energy theory is a gauge theory governed by a free
infrared (IR) fixed point such as an Abelian gauge theory with
massless fermions or a non-Abelian gauge theory with many massless
flavors~\cite{Appelquist:1999hr}, where the $\beta$ function is
positive at weak coupling and $\langle F^2 \rangle$ 
is known to be negative, then the trace
anomaly~\eqref{eq:ta} becomes negative.
We also point out that some phenomenological nuclear EoSs bear a
negative trace anomaly due to sudden stiffening of the EoS with
$P > \varepsilon/3$~\cite{Zeldovich:1961sbr,Serot:1997xg,Akmal:1998cf}.
Moreover, QCD at finite isospin chemical
potential~\cite{Son:2000xc,*Son:2000by} and two-color QCD at finite
$\muB$~\cite{Cotter:2012mb, Iida:2022hyy} produce a negative trace anomaly.

Nevertheless, in view of the observational data in
Fig.~\ref{fig:NS}, QCD may well enjoy a special property that
the matter part of the trace anomaly is positive definite.
One supportive argument is based on the behavior of the
chromoelectric field, $\boldsymbol{E}$, and the chromomagnetic field,
$\boldsymbol{B}$.  In the chiral limit only the gluon condensate,
$\langle F^2 \rangle_{\muB}
= \langle \boldsymbol{B}^2 - \boldsymbol{E}^2 \rangle_{\muB}$,
contributes to the trace anomaly.
Nuclear matter at low density is approximated as a gas of nucleons,
and the trace anomaly is positive for each nucleon (that is the
nucleon mass squared), and so the trace anomaly in dilute nuclear
matter should be positive.  In the nonrelativistic quark model at
higher density, the interquark interaction is dominantly mediated by
the chromoelectric field, and so the trace anomaly is positive.
Besides, we know for sure that the direct pQCD computation at
asymptotic high density gives a positive trace anomaly.

From another perspective the positivity of the trace
anomaly can be motivated as follows.
Eq.~\eqref{eq:tracedof} relates the matter part of the trace anomaly
to the density derivative of effective degrees of freedom $\nu_\mu$.
As long as more effective degrees of freedom 
are liberated at higher $\muB$,
we can conclude $\langle \tam \rangle_{\muB} \geq 0$ because of
$d\nu_\mu / d\muB \geq 0$.
It is an intriguing question how the above argument could be modified
if color superconductivity is activated with a finite condensation of
quark Cooper pairs.

To prove $\langle \tam \rangle_{\muB} \geq 0$ directly from QCD
is an intriguing challenge.  It is a nontrivial and profound
question due to the composite operator renormalization.  Here, we
propose a complementary strategy to test this conjectured
inequality using astronomical observations of NSs, namely,
the maximum mass bound.

One-to-one correspondence is established between the EoS and $M(R)$
(where $M$ is the NS mass as a function of the NS
radius $R$).
In order to find the maximum mass,
$M_{\text{max}}(R)$, for a given radius $R$.
we assume a standard crust EoS up to $\nB \leq 0.5 n_0$~\cite{Baym:1971pw,Negele:1971vb}.
Then, for $\nB > 0.5 n_0$ we identify $M_{\text{max}}(R)$ by taking
the maximally stiff/soft EoS parametrizations.
Technical details are outlined in Refs.~\cite{Rhoades:1974fn, Koranda:1996jm} (see also Ref.~\cite{Drischler:2020fvz}).

\begin{figure}
  \includegraphics[width=\columnwidth]{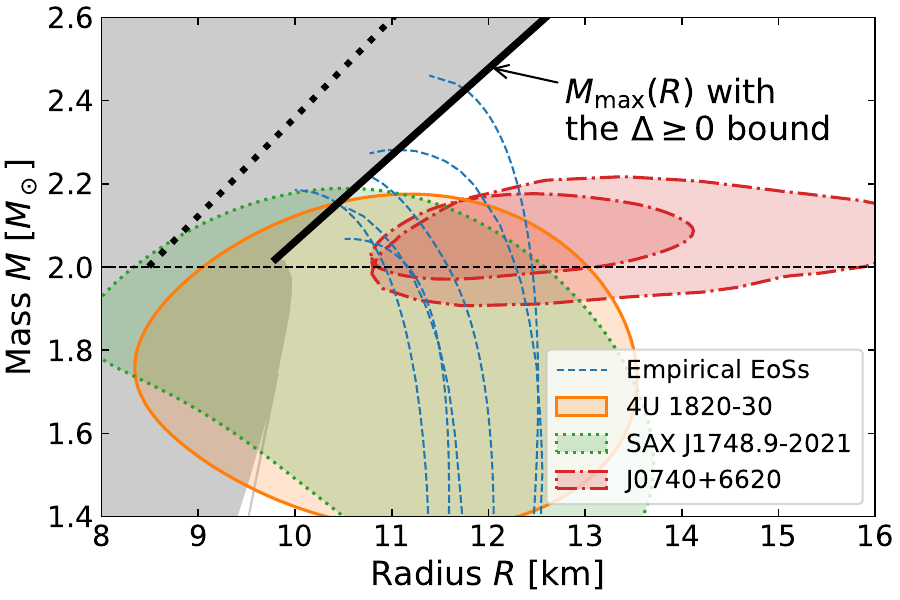}
  \caption{The effect of the $\Delta \geq 0$ bound on the NS $M$-$R$
    relation.  The black solid
    (dotted) line shows the maximum mass configuration for the EoS
    with (without) the $\Delta \geq 0$ bound.  We also overlay the
    measurement of NSs and the $M$-$R$ relations (thin dashed curves)
    corresponding to
    empirical nuclear EoSs from Refs.~\cite{Akmal:1998cf, Goriely:2010bm,
      Engvik:1995gn, Baym:2019iky} and two variants
    from Ref.~\cite{Muther:1987xaa}\footnote{The data
      are adopted from \texttt{http://xtreme.as.arizona.edu/NeutronStars/}}.}
  \label{fig:MR}
\end{figure}

Some maximally stiff EoS may render negative $\Delta$.
In Fig.~\ref{fig:MR} the dotted line represents the original
$M_{\text{max}}(R)$, while
the black solid line shows $M_{\text{max}}(R)$
for the EoS with the $\Delta \geq 0$ condition taken into account.
Performing the EoS scan we find the gray shaded region that is
incompatible with the $\Delta \geq 0$ condition.
For completeness we overlay three current radius measurements obtained
with two different methods; namely, spectral measurement of 4U 1820-30
and SAX J1748.9-2021~\cite{Ozel:2015fia}, as well as the timing
measurement of J0740+6620 from
NICER~\cite{Riley:2021pdl,Miller:2021qha}.
We also plot the $M$-$R$ relations from empirical nuclear
EoSs~\cite{Akmal:1998cf, Goriely:2010bm, Engvik:1995gn,
  Muther:1987xaa, Baym:2019iky}.
From Fig.~\ref{fig:MR} we can say that the $\Delta \geq 0$ condition
has a phenomenological impact to tighten the allowed $M$-$R$ region.
In Fig.~\ref{fig:MR} we put a thin line at $M/M_{\odot}=2$ for
eye guide.  If the maximum mass is larger as reported in
Refs.~\cite{Romani:2022jhd, Linares:2018ppq}, our proposed bound
would exclude EoSs that lead to sufficiently heavy mass but small
$R$ inside the gray shaded region.
We propose further systematic comparisons of results
with/without our positivity condition as well as the hypothesised
conformality bound on the sound velocity for other observables such
as the tidal deformability along the lines of, e.g.,
Refs.~\cite{Annala:2017llu,Annala:2021gom}.
Future multimessenger observations, which are expected to pin down the
maximum mass of
NSs~\cite{Margalit:2017dij,Shibata:2017xdx,Rezzolla:2017aly,Ruiz:2017due},
and radius measurements together with the tidal deformability
  inferred from the merger will help to test our conjecture of the positive trace
anomaly.
\vspace{0.5em}

\begin{acknowledgments}
  We thank Neill~Warrington for discussions.
  Y.~F.\ would like to acknowledge useful conversations with
  Greg~Jackson and Sanjay~Reddy.
  K.~F.\ thanks Shi~Chen for illuminating discussions.
  The work of Y.~F., L.~M., and M.~P.\ was supported by the U.S. DOE
  under Grant No. DE-FG02-00ER41132.
  K.~F.\ was supported by JSPS KAKENHI Grant Nos. 22H01216 and 22H05118.
\end{acknowledgments}

\bibliographystyle{apsrev4-1}
\bibliography{bib_trace}

\end{document}